# 클라우드 환경에서 수사 실무와 법적 과제

# Practical and Legal Challenges of Cloud Investigations

조슈아 제임스[*], 장윤식[**]

**Joshua I. James[*], Yunsik Jang[**]**

요 약 클라우드 컴퓨팅 서비스의 확산으로 범죄수사를 위한 증거수집의 관점에서 불확실성으로 인한 다양한 실무적이고 법적인 문제가 제기되고 있다. 이 논문은 클라우드 환경에 대한 일반적인 수사상의 논점을 개관하고, 관할과 국제공조를 비롯한 문제점을 진단한다. 실무적으로 직접적으로 수사관이 접속하는 경우와 서비스제공자의 협조를 받는 경우의 장단점을 비교하여 실무적 개선방안을 논의하고 이에 따른 관할의 중복과 서비스 약정 및 포렌식적으로 무결한 데이터 수집 등 법률적 쟁점을 정리한다.

**Abstract** An area presenting new opportunities for both legitimate business, as well as criminal organizations, is Cloud computing. This work gives a strong background in current digital forensic science, as well as a basic understanding of the goal of Law Enforcement when conducting digital forensic investigations. These concepts are then applied to digital forensic investigation of cloud environments in both theory and practice, and supplemented with current literature on the subject. Finally, legal challenges with digital forensic investigations in cloud environments are discussed.

## Ⅰ. Cloud Computing and Digital Forensic Investigation

As more businesses and end-users adopt technologies that utilize cloud technologies[1], criminals too begin to use and exploit such technologies. Because of the complexity in setup, usage, hosting and even location of cloud services, digital forensic investigation of such technologies can be challenging. A number of prior works have looked at the challenges of extracting digital evidence from cloud environments [2][3], in this work we will provide an overview of the state of Cloud investigation practice and theory and describe legal challenges that are commonly observed by Law Enforcement.

The job of police officers – digital forensic investigators – is normally to conduct an impartial investigation, and provide an unbiased report on the discovered facts – both inculpatory and exculpatory – that are relevant to the probandum. Discovered facts can be submitted as evidence, which are then tested by the courts to determine if the evidence is admissible. Evidence from digital forensic investigations is considered scientific, like other forensic sciences, and therefore can be tested against the requirements for the admittance of scientific evidence. Assuming the evidence is admitted in court, a jury then compares the observed chain of events to the probandum, and guilt or innocence is determined.

[*]정회원, 순천향대학교 법과학대학원
[**]정회원, 순천향대학교 법과학대학원(교신저자)
접수일자: 2014년 11월 1일, 수정일자: 2014년 12월 1일
게재확정일자: 2014년 12월 12일

Digital evidence is defined as "[i]nformation of probative value that is stored or transmitted in binary form".[4] There are two states of data that digital forensic investigators must work with: live data and persistent (post-mortem) data. Live data is data in a system that is powered on (like in the case of Cloud hosts and instances), and is more prone to change. Persistent data is data available when the system has been shut down. It must be said that all data has a degree of volatility, or susceptibility to change, and the speed in which data is likely to change, ordered from fastest to slowest, is known as the Order of Volatility (OoV).

Digital investigators must consider the implications of collecting each type of data, and be able to prioritize the data by the order of volatility. The order of volatility, and even availability of data, differs between a live system and an offline system. A number of prior works have discussed specific challenges with the collection, verification and preservation of digital evidence from Cloud environments.[5][6]

Sources of digital evidence are dependent on the current state of the suspect system. For example, if a suspect's system is live, data from RAM may be accessible. This data could possibly hold information that could be used as inculpatory or exculpatory evidence, but would not be available if the system was powered down. While computer systems are common sources of digital evidence, other digital devices, such as cellular phones, are becoming just, if not more, common.

## II. Digital Forensic Investigation in Cloud Environments

New concepts in cloud computing have created new challenges for security teams and researchers alike. Cloud computing service and deployment models have a number of potential benefits for businesses and customers, but security and investigation challenges – some inherited from 'traditional' computing, and some unique to cloud computing – create uncertainty and potential for abuse as cloud technologies proliferate.

### 1. Cloud Computing and Investigation Challenges

A 2013 survey [7] of 106 participants identified the main challenges of cloud investigation as:

  a. Jurisdiction

  b. Lack of international collaboration and legislative mechanism in cross-nation data access and exchange

  c. Lack of law/regulation and law advisory

  d. Simple role management (e.g., admin, user) makes it difficult to categorize suspects

  e. Investigating external chain of dependencies of the cloud provider (e.g., a cloud provider can use the service from another provider)

  f. Decreased access to and control over forensic data at all levels from customer side

  g. Exponential increase of digital (mobile) devices accessing the cloud

Many of the technologies that make up cloud services, such as virtualization, have existed since before the utility computing business model was practical on a large scale. Because infrastructure, platform and software hosting technologies have existed for business and personal use for some time, techniques for digital investigation of incidents relating to these technologies may be well known. For example, forensic acquisition and verification of hard drives is a common task in digital forensic investigations. Many times the same acquisition methods for a physical disk may also work for virtual disks associated with a virtual machine. Many of the digital forensic acquisition, verification and analysis techniques may be able to be applied to cloud investigations, but cloud does pose some new challenges for digital forensic investigators. The following are a few of the many potential challenges cloud computing brings to digital investigation.

In traditional investigations the suspect or victim's computers may normally be the main source of information about an incident. For example, in a child exploitation case, a suspect may have stored illicit images on their local hard drive. By finding and analyzing the images, the investigator may be able to determine that the images were, in fact, illegal. Other information, such as the recently opened files list, may be used to support that the suspect had knowledge of the images. However, if the suspect is using cloud-based storage, the images may not be stored locally. In this case, the investigator may be able to show that the suspect had knowledge of the images, but not whether the images were actually illegal. Saliba(2012) claims that cloud services may reduce the amount of direct evidence available on a suspect's disk[8], but sometimes provide more information about the user and cloud service that would help in acquiring a subpoena or warrant. As more data is stored in the cloud, and services reduce the client-side impact, fewer evidential traces may be found.

When suspect data is stored in the cloud the data may be in one jurisdiction while the suspect machine connecting to the service may be in another jurisdiction. This scenario is becoming a challenge for Law Enforcement in many countries, especially with mobile cloud-attached devices. In many countries, investigators may access data stored remotely if given permission by the suspect (verbal conformation, entering the password, etc.). However, there has been very little definition of what to do if data is stored on a non-national cloud service that is currently connected while the investigator begins a live analysis of the suspect system. If the data is accessible, an investigator may save a considerable amount of time by acquiring the data from the connected service rather than waiting for international requests. But authority on this matter is not always clear. A lack of definition on the scope of acquisition of data on non-national remote connections sometimes depends on the country, and many times depends on the investigator's preliminary analysis of the remotely stored data as well as the likelihood of receiving the data if an international request was made.

Investigators physically accessing the CSP may also be more difficult, and may be impossible if the data is distributed over several geographic locations.[9] In traditional investigations, a computer or server may possibly be taken down, and its physical disks imaged. Taking down production servers is increasingly rare for server environments since law enforcement may be liable for damages while the server is down. Taking servers down in a cloud environment may have an impact on many customers, creating more liability. Further, data may be stored on virtual storage spanning multiple servers, or even geographic locations, meaning that hard disk acquisition may not be practical, or even produce the desired data. Liability and reconstruction of virtual storage in cloud environments from physical disk images remains a challenge.

Storage in the cloud is attractive to users because it is highly accessible and relatively inexpensive. As previously mentioned, the amount of data on personal hard drives is becoming too large for most law enforcement agencies to acquire and store all the data. Cloud services, however, are currently offering Gigabytes of space for free with the option to pay for more storage. The amount of stored data could quickly add up across multiple cloud service offerings. Resulting again in too much data for law enforcement to process and store.

Since acquisition of a whole physical disk may not be practical or possible, and the quantity of data may be too large to effectively process and store, selective data acquisition may be required. Selective data acquisition implies a preliminary analysis, or some prior knowledge, to reduce the overall dataset an investigator is interested in. The challenge with this method is an intrinsic challenge in digital forensic investigations; how do we know what we don't know? In other words, even if all possible data could be acquired, how do we know that no evidence has been missed? If this question cannot be easily answered when all data is available, how can an investigator justify reducing the dataset, and potentially excluding inculpatory and/or exculpatory evidence? Some investigators are currently focusing on data sources that they believe are likely to provide the richest sources of information, but justifiable exclusion remains a challenge.

Because of the distributed, multi-layered nature of cloud computing, chain of custody for the data may be impossible to verify. Without strict controls it may be impossible to determine where exactly the data was stored, who had access, and was leakage or contamination of data possible. If data is stored in a cloud where multiple users and CSPs potentially have access, association of the data to the suspect is a challenge to establish beyond a reasonable doubt.

When an incident occurs on the side of the CSP, the CSP may be more concerned with restoring service than with preserving evidence. Further, the CSP may begin its own investigation into an incident without taking proper precautions to ensure the integrity of potential evidence. In more severe cases, CSPs may not report or cooperate in investigation of incidents for fear of reputational damage. The challenge in this case is with the competence and trustworthiness of the CSP. A CSP would be an effective, immediate first-responder, but questions about the integrity and chain of custody of the acquired evidence may make admissibility difficult. To meet this challenge law enforcement should work with CSPs, and ensure proper documentation is being created and forensically sound processes are being used.

Another challenge is with feature of rapid elasticity in cloud environments. Data associated with newly created virtual machine instances may only be available for a limited time. To this author's knowledge, no research has been conducted on determining available data associated with removed VM instances. If a new VM instance is created and either compromised or used to attack, evidential traces may be available in the VM. If the VM instance is then deallocated, investigators currently do not know whether evidential traces, or the entire VM instance cloud be recovered.

The final challenge in this non-comprehensive list is the issue of international communication. As mentioned previously, cloud computing blurs physical, policy and jurisdictional boundaries globally. However, law enforcement at a global level has yet to find effective, timely and efficient international communication and cooperation channels. Conferences such as the International Symposium on Cybercrime Response specifically discuss international law enforcement communication and collaboration efforts. Such conferences allow law enforcement to create informal communication channels, and sometimes help in the creation of bilateral agreements for cooperation, but these channels have their limits.

Global law enforcement communication channels, such as INTERPOL's I-24/7 network or the G8 24/7 network, connect many countries, but are limited by their structure and bureaucracy. Many officers have found the global networks to be somewhat effective if the request was not overly urgent, however, these networks have failed to address real-time requests for help from countries under DDoS attack. Many times, law enforcement will prefer faster, informal channels to begin an international investigation, rather that traversing such networks. However, multi-country operations, such as 'Operation Unmask', show the potential of these networks to assist in large-scale coordination efforts. Overall, users, businesses and even criminals are utilizing technologies, such as cloud computing, to be able to rapidly find and share (and exploit) new ideas. These groups are no longer considering physical and political borders. Law enforcement, however, is currently restricted by a lack of effective global communication channels, political issues, and jurisdiction that make policing in a globally connected world even more of a challenge.

## Ⅲ. Cloud Forensics in Practice

In practice, digital investigations in cloud environments are taking place largely as before. When attempting to access evidential traces on cloud-connected devices, 'traditional' digital forensic investigation methods can be employed, such as computer or mobile device analysis. The greater challenge lies in the fact that a large amount of data only be available in the Cloud. In practice there are essentially two options that investigators have to access data stored in the Cloud. They can either attempt to access the data themselves through user authentication methods, or work with the Cloud Service Provider to collect data.

Normally, attempting to access the data directly is preferred because it is much faster than making formal requests. Further, there is no guarantee that a CSP will comply with a request for data, especially if the CSP is outside of the jurisdiction of the investigator making the request.

Many times, however, the local device, such as the suspect's computer or phone will contain authentication data that may allow the investigator to access data stored on the cloud. There is also the possibility that the suspect willingly supplies such credentials. If the investigator can get access to suspect data stored in the Cloud, the cloud service will determine whether the investigator can use standard forensic acquisition tools (Dykstra and Sherman 2012).[10]

For example, investigators may access cloud-based storage and attempt to acquire suspect data directly from the logical device. In many cases, an investigator may be forced to used the interface provided by the CSP. Forensic file copying is preferred over acquiring an image of the logical volume, even though investigators may not always understand when (and how) file content and meta- data are modified by the cloud service. Acquisition of a physical disk that is part of a cloud environment is rarely, if ever, done because of the size of physical and reconstructing data is not always feasible.

If the investigator cannot access the data directly – either technically or legally – then participation from the Cloud Service Provider may be necessary. In this case, the type of data that the CSP can, or is willing to, provide the investigator will depend on a number of factors including the CSP's internal policies, jurisdiction and general attitude toward the investigator/case. If data is provided it could be anything from logs relating to the suspect to file content. Again, the type, quality and quantity of available data can vary greatly between CSP.

### 1. Opportunities for Digital Investigations in Cloud Environments

Ruan, Carthy et al. (2011) also identified a number of potential opportunities that cloud environments bring to digital forensic investigations.[11] The first is an improvement of cost effectiveness when implementing 'forensic services' on a large scale. In this case, it is assumed that forensic services that are processing-intensive can also using cloud infrastructure to deploy forensic services and offer such services to the masses.

The next opportunity is the possibility of a greater amount of data being made available to an investigator. Multiple copies of data may be stored in a cloud service. If one copy is deleted, damaged or otherwise altered an investigator may be able to access backup copies of the data. Further, cloud services often include comprehensive logs of transactions, which may also help in investigations. Further, cloud services may help investigations with processing and data validation since cloud services normally integrate data validation, and are tuned for data processing tasks.

## Ⅳ. Legal Challenges Relating to Cloud Investigations

Commonly discussed legal concerns with cloud investigations include issues with jurisdiction, international law and cooperation; Service Level Agreements; and forensic data collection.

### 1. Multiple Jurisdictions

Jurisdiction is a concern in cloud environments because a CSP may be located in several different legal jurisdictions. The data, then, may be subject to multiple legal considerations at the same time. The situation is made even more complex when one CSP is using services from another CSP that may be located in separate jurisdictions.

An investigator may not have legal authority to acquire data that is outside of their jurisdiction, even if that data can be easily accessed. Multiple jurisdictions normally mean that an investigator will need to make a mutual legal assistance request for data, which will be forwarded to an agent in the jurisdiction of interest for processing. Formal mutual legal assistance requests normally take a long time. This combined with the time the CSP may take to respond to the request could mean months until the investigator receives data, if at all. Overall, data resident in multiple jurisdictions is a challenge because of politics. International cooperation in digital investigations works, but not well. Some regions work better with each other than others. Challenges are mostly due to communication and sovereignty issues that result in long delays or no response to requests. Technically multiple jurisdictions pose few new challenges since cloud services are hyper-connected, it is technically easy to transfer data. However, it is difficult for an investigator to get the authority to do so in another jurisdiction.

### 2. Service Level Agreements

The need for establishing Service Level Agreements (SLA) considering digital investigations and security in cloud environments has been widely discussed. In many of these works, the authors claim the SLA should, most of all, focus on who is responsible for which security/ investigation/ information sharing tasks, and when and how these tasks should be carried out. The SLA should also ensure that the CSP can comply with applicable regulations and industry standards. The customer should be made aware of which laws (which jurisdiction) govern the SLA.

### 3. Forensic Data Collection

Forensic data collection is another legal challenge relating to the acceptance of digital evidence in court. An investigator should be able to verify the data they are basing any conclusions on accurately represents the suspect's original data. The difficulty for investigators is understanding how digital evidence can potentially change based on the service that was used. For example files stored with Amazon and Google cloud services may alter timestamps, or even file content, differently. Beyond the services, it is not always clear how forensic tools will alter data stored on these services. Research is currently looking into these questions, but it is difficult when so many different cloud offerings exist with potentially different standards.

Another challenge related to forensic data collection is the challenge of evidence segregation. Cloud services potentially create a large amount of data. However, data, such as logs, may contain information about multiple users. Further, if infrastructure is shared between multiple users, it is difficult to segregate data for a user of interest.

Finally, external dependency chains further complicate the issue of trust in the CSP to accurately represent collected data. CSPs and their dependents should be able to demonstrate that integrity has been maintained, but ultimately investigators, and the legal system in general, will have to accept some level of risk associated with CSP and their dependents providing data that will be used as evidence.

## Ⅴ. Conclusions

Digital forensic investigations of cloud environments is a complicated topic that covers technical, organizational and legal considerations. While digital forensic investigations are taking place in cloud environments using traditional investigation methods, there are still many questions. These questions will evolve as cloud computing continues to evolve, however, the legal challenges will largely remain the same. Namely, the question of jurisdiction and international cooperation.

저자 소개

### Joshua Issac James(정회원)

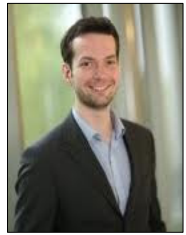

- 2013년 9월 : University College Dublin, Dublin, Ireland. PhD Computer Science by research in Digital Forensic Investigation
- 2014년 ~ 현재 : 순천향대학교 법과학대학원 초빙교수
- Joshua@CybercrimeTech.com

### 장 윤 식(정회원)

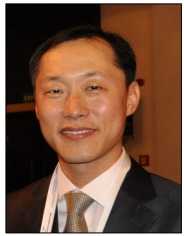

- 2001년 2월 : 고려대 법무대학원 졸업 (법학석사)
- 2014년 8월 : 고려대 정보경영공학전문대학원 졸업(공학박사)
- 2014년 ~ 현재 : 순천향대학교 법과학대학원 초빙교수
- ccismem@gmail.com